\documentclass
[a4paper,aps,prd,twocolumn,preprintnumbers,showpacs,superscriptaddress,nofootinbib]{revtex4}%
\usepackage{graphics,graphicx,epsfig}
\usepackage{amsfonts,amsmath,amssymb}
\usepackage{amsmath}
\usepackage{amsfonts}
\usepackage{amssymb}
\usepackage{color}
\usepackage{graphicx}%
\usepackage{epstopdf}
\begin{document}
\title{The different phases of hairy black holes in AdS$_{5}$ space}
\author{Gaston Giribet}
\email{gaston-at-df.uba.ar}
\affiliation{Departamento de F\'{\i}sica, Universidad de Buenos Aires and IFIBA-CONICET,
Ciudad Universitaria, Pabell\'on I, 1428, Buenos Aires, Argentina.}
\affiliation{Instituto de F\'{\i}sica, Pontificia Universidad Cat\'{o}lica de
Valpara\'{\i}so, Casilla 4059, Valpara\'{\i}so, Chile.}
\author{Andr\'es Goya}
\email{af.goya-at-df.uba.ar}
\affiliation{Departamento de F\'{\i}sica, Universidad de Buenos Aires and IFIBA-CONICET,
Ciudad Universitaria, Pabell\'on I, 1428, Buenos Aires, Argentina.}
\author{Julio Oliva}
\email{julio.oliva-at-uach.cl}
\affiliation{Instituto de F\'{\i}sica y Matem\'{a}tica, Universidad de Ciencias,
Universidad Austral de Chile UACh, Valdivia, Chile.}

\pacs{Pacs: 42}

\begin{abstract}
We investigate the thermodynamics of hairy black holes in asymptotically
Anti-de Sitter (AdS) space, including backreaction. Resorting to the Euclidean
path integral approach, we show that matter conformally coupled to Einstein
gravity in five dimensions may exhibit a phase transition whose endpoint turns out to be a hairy black hole in
AdS$_{5}$ space. The scalar field configuration happens to be regular
everywhere outside and on the horizon, and behaves asymptotically in such a
way that respects the AdS boundary conditions that are relevant for AdS/CFT.
The theory presents other peculiar features in the ultraviolet, like the existence of black holes
with arbitrarily low temperature in AdS$_{5}$. This provides a simple setup
in which the fully backreacting problem of a hair forming in AdS at certain critical temperature can be solved analytically.

\end{abstract}
\maketitle

\section{Introduction}

According to AdS/CFT correspondence \cite{Maldacena}, asymptotically
five-dimensional Anti-de Sitter (AdS$_{5}$) black holes describe the dynamics
of finite temperature phases of four dimensional gauge theories \cite{Witten}.
Phenomena such as confinement, thermalization, and phase transitions are all
described holographically in terms of classical black hole geometries in AdS
space. In the study of holographic superconductors, for instance, the gravity
dual description is that of a scalar field condensating around an
asymptotically AdS black hole at sufficiently low temperature \cite{HHH,
Gubser}. When forming in the bulk, the scalar hair induces the expectation
value of operators in the dual field theory.

The relation between bulk and boundary phases makes that solving the
backreacting problem in the AdS black hole background analytically is of big
interest, as it would provide a detailed description of finite temperature
regimes of strongly coupled systems and, in particular, their phase
transitions. Unfortunately, solving the problem taking into account
backreaction is highly non-trivial, and the models in which one can actually
do so explicitly are limited and hard to come up with. However, a notably
simple scenario in which the problem can be solved analytically is provided by
general relativity (GR) coupled to conformally coupled scalar field matter. In
fact, this model does admit hairy black holes in asymptotically flat and
asymptotically (A)dS spaces in arbitrary dimensions $D>4$, \cite{nosotros}.
The scalar field configuration of these hairy black holes is regular
everywhere outside and on the horizon, and in the case of negative
cosmological constant it behaves asymptotically in a manner that respects
the AdS boundary conditions relevant for AdS/CFT. In this paper, we will study
the thermodynamics of these hairy black hole solutions and, in particular,
their phase transition. We will show that, at high temperature, black holes
with non-vanishing scalar hair configurations are thermodynamically favored
with respect to the non-hairy Schwarzschild-AdS$_{5}$ solution and thermal
AdS$_{5}$. Below certain critical temperature, the theory undergoes a phase
transition of the Hawking-Page type \cite{HawkingPage} and the dominant
configuration happens to be thermal AdS$_{5}$. The presence of the scalar
conformal field also changes the thermodynamics of the theory at short
distance:\ In particular, we will show that, unlike what happens in GR in
absence of matter, small black holes with conformally coupled scalar hair with
arbitrarily low temperature exist in AdS$_{5}$. These small black holes
present a positive specific heat, and, as a consequence of that, become stable
under Hawking radiation, yielding a remnant with finite mass. However, these
small black holes are not stable under tunneling transitions; for the range of
temperatures they exhibit, the free energy of thermal AdS$_{5}$ is actually lower.

The paper is organized as follows:\ In Section II we introduce the theory,
consisting of scalar field matter conformally coupled to GR in five
dimensions. In Section III we review the main properties of the hairy black
hole solutions in AdS$_{5}$ space. In Section IV we compute the mass of these
black holes using the Hamiltonian formalism, which we later confirm by using
Euclidean action methods. In Section V, we study the thermodynamics of
AdS$_{5}$ black holes including the effects of the scalar field hair. In
Section VI, we consider the Euclidean action in the saddle point approximation
and derive the mass, the entropy, and the free energy of the hairy black hole
solutions. This allows us to study, in Section VII, the phase transitions
between different configurations that coexist at different ranges of
temperature. We present our conclusions in Section VIII.

\section{The theory}

In $D$ space-time dimensions we consider the action%
\begin{equation}
\mathcal{I}=\frac{1}{\kappa}\int d^{D}x\sqrt{-g}\left(  R-2\Lambda
+\kappa\mathcal{L}\left(  \phi,\nabla\phi\right)  \right)  \label{I}%
\end{equation}
where $\kappa=16\pi G$ and where $\mathcal{L}\left(  \phi,\nabla\phi\right)  $
represents the Lagrangian of a conformally coupled real scalar field. Apart
from the requirement of conformal invariance, the matter Lagrangian is chosen
in a way it yields second order field equations. Provided these requirements,
$\mathcal{L}\left(  \phi,\nabla\phi\right)  $ can be identified uniquely
\cite{weones}.

To express the conformal matter content in a succinct way, it is convenient to
introduce the four-rank tensor\footnote{This corresponds to considering the
choice $s=-1/3$ in Eq. (2) of Ref. \cite{weones}.}%
\begin{align}
S_{\mu\nu}^{\quad\gamma\delta}  &  =\phi^{2}R_{\mu\nu}^{\quad\gamma\delta
}-12\delta_{\lbrack\mu}^{[\gamma}\delta_{\nu]}^{\delta]}\nabla_{\rho}%
\phi\nabla^{\rho}\phi-\nonumber\\
&  48\phi\delta_{\lbrack\mu}^{[\gamma}\nabla_{\nu]}\nabla^{\delta]}%
\phi+18\delta_{\lbrack\mu}^{[\gamma}\nabla_{\nu]}\phi\nabla^{\delta]}%
\phi\label{Sij}%
\end{align}
which can be shown to transform covariantly under Weyl rescaling. More
precisely, if one performs the Weyl transformation%
\begin{equation}
g_{\mu\nu}\rightarrow\Omega^{2}g_{\mu\nu},\qquad\phi\rightarrow\Omega
^{-1/3}\phi
\end{equation}
tensor (\ref{Sij}) transforms as%
\begin{equation}
S_{\mu\nu}^{\quad\gamma\delta}\rightarrow\Omega^{-8/3}S_{\mu\nu}^{\quad
\gamma\delta}.
\end{equation}

In terms of this tensor, the matter Lagranian $\mathcal{L}(\phi,\nabla\phi)$
takes the form%
\begin{align}
\mathcal{L}(\phi,\nabla\phi)  &  =\sum_{k=0}^{\left[  \frac{D-1}{2}\right]
}b_{k}\frac{k!}{2^{k}}\ \phi^{3D-8k}\delta_{\lbrack\alpha_{1}}^{\mu_{1}}%
\delta_{\beta_{1}}^{\nu_{1}}...\delta_{\alpha_{k}}^{\mu_{k}}\delta_{\beta
_{k}]}^{\nu_{k}}\ \times\nonumber\\
&  \qquad\times S_{\quad\mu_{1}\nu_{1}}^{\alpha_{1}\beta_{1}}...S_{\quad
\mu_{k}\nu_{k}}^{\alpha_{k}\beta_{k}}\,, \label{L}%
\end{align}
where the symbol $[n]$ stands for the integer part of $n$.

In $D=4$ dimensions, after the field redefinition $\phi\rightarrow\phi^{1/3}$, Lagrangian (\ref{L}) corresponds to the conformally
coupled matter consisting of a canonically normalized scalar field coupled to
the spacetime curvature with a term $-\frac{1}{12}R\phi^{2}$ and with a
$\lambda\phi^{4}$ potential; see \cite{weones, nosotros}\ for details. In
$D=5$, which is the case we are interested in here, Lagrangian (\ref{L}) takes
the form
\begin{equation}
\mathcal{L}(\phi,\nabla\phi)=\phi^{15}\left(  b_{0}\mathcal{S}^{(0)}+b_{1}%
\phi^{-8}\mathcal{S}^{(1)}+b_{2}\phi^{-16}\mathcal{S}^{(2)}\right)  \label{np}%
\end{equation}
where%
\begin{align*}
\mathcal{S}^{(0)}  &  =1,\\
\mathcal{S}^{(1)}  &  =S\equiv g^{\mu\nu}S_{\mu\nu}\equiv g^{\mu\nu}%
\delta_{\sigma}^{\rho}S_{\ \mu\rho\nu}^{\sigma},\\
\mathcal{S}^{(2)}  &  =S_{\mu\nu\alpha\beta}S^{\mu\nu\alpha\beta}-4S_{\mu\nu
}S^{\mu\nu}+S^{2}.
\end{align*}
This contains higher-curvature couplings, and such couplings are actually the
reason why this model happens to circumvent no-hair theorems \cite{nosotros}.
Notice that all terms in Lagrangian (\ref{np}) are well-behaved in the limit
$\phi$ going to zero (including the third one) since the tensor $S_{\mu\nu
}^{\quad\gamma\delta}$ is quadratic in the scalar field.

Field equations coming from action (\ref{I}) are Einstein equations%
\begin{equation}
R_{\mu\nu}-\frac{1}{2}Rg_{\mu\nu}+\Lambda g_{\mu\nu}=\kappa T_{\mu\nu
}\ \label{GT}%
\end{equation}
with the energy-momentum tensor%
\begin{align}
T_{\mu}^{\nu}  &  =\sum_{k=0}^{\left[  \frac{D-1}{2}\right]  }\frac{k!b_{k}%
}{2^{k+1}}\phi^{3D-8k}\delta_{\lbrack\mu}^{\nu}\delta_{\rho_{1}}^{\lambda_{1}%
}...\delta_{\rho_{2k}]}^{\lambda_{2k}}\times\nonumber\\
&  \qquad\times\ S_{\ \ \ \ \lambda_{1}\lambda_{2}}^{\rho_{1}\rho_{2}%
}...S_{\ \ \ \ \lambda_{2k-1}\lambda_{2k}}^{\rho_{2k-1}\rho_{2k}}\ ,
\label{TTT}%
\end{align}
and the equation for the scalar field%
\begin{align}
0  &  =\sum_{k=0}^{\left[  \frac{D-1}{2}\right]  }\frac{\left(  D-2k\right)
k!b_{k}}{2^{k}}\phi^{3D-8k-1}\delta_{\lbrack\alpha_{1}}^{\mu_{1}}\delta
_{\beta_{1}}^{\nu_{1}}...\delta_{\alpha_{k}}^{\mu_{k}}\delta_{\beta_{k}]}%
^{\nu_{k}}\times\nonumber\\
&  \qquad\times\ S_{\quad\mu_{1}\nu_{1}}^{\alpha_{1}\beta_{1}}...S_{\quad
\mu_{k}\nu_{k}}^{\alpha_{k}\beta_{k}}. \label{eqfieldarb}%
\end{align}

Equations (\ref{GT})-(\ref{eqfieldarb}) are of second order both in the metric
and in the scalar field. When these equations are imposed, the trace of
energy-momentum tensor (\ref{TTT}) of course vanishes.

\section{Black hole solution}

The theory defined by action (\ref{I}) admits analytic black hole solutions
with non-vanishing scalar field \cite{nosotros}. These solutions co-exist with
those of GR, that also solve (\ref{GT})-(\ref{eqfieldarb})\ with $\phi=0$. The
hairy ($\phi\neq0$)\ black hole solutions, which admit horizons of diverse
topology, are known analytically and present interesting properties,
especially a very interesting thermodynamics. The scalar field configuration
is regular everywhere outside and on the horizon, and backreacts on the metric
in such a way that the mass of the black hole remains finite. Moreover, the
scalar field and the deformation of the metric due to its backreaction vanish
at large distance sufficiently fast for the (A)dS asymptotics not to result distorted.

The metric of these black hole solutions take the form
\begin{equation}
ds^{2}=-N^{2}(r)f(r)\ dt^{2}+\frac{dr^{2}}{f(r)}+r^{2}d\Omega_{3}^{2}
\label{g1}%
\end{equation}
with
\begin{equation}
f(r)=1-\frac{m}{r^{2}}-\frac{q}{r^{3}}-\frac{\Lambda}{6}r^{2},\qquad
N^{2}(r)=1; \label{g2}%
\end{equation}
and where $d\Omega_{3}^{2}$ is the metric of the unit \thinspace$3$-sphere,
which has volume $V=2\pi^{2}$. Here, we will be concerned with the case
$\Lambda<0$. Solution (\ref{g1})-(\ref{g2}) shares some features with the
lower-dimensional hairy black holes\footnote{First order phase transitions for four-dimensional black holes with a minimally coupled self-interacting scalar hair has been recently found in \cite{choqueetal}.} \cite{MTZ2}-\cite{Astorino}.

The scalar hair configuration is%
\begin{equation}
\phi(r)=\dfrac{n}{r^{1/3}} \label{phi}%
\end{equation}
which is regular everywhere but at the origin. The scalar field backreacts on
the geometry through the term $\sim1/r^{3}$ in (\ref{g2}), which, as said,
does not spoil the AdS$_{5}$ asymptotic behavior of the metric near the boundary.

Parameter $m$ in (\ref{g2})\ is an arbitrary integration constant, while $n$
in (\ref{phi}) and $q$ in (\ref{g2}) are fixed in terms of the couplings
$b_{i}$ ($i=0,1,2$) by the following relations
\begin{equation}
q=\frac{64\pi G}{5}b_1n^{9}\,\ \ \ \ \ \ ,\quad n=\varepsilon
\left(  -\frac{18}{5}\frac{b_{1}}{b_{0}}\right)  ^{1/6}, \label{conditions}%
\end{equation}
where $\varepsilon=-1,0,+1$. Conditions (\ref{conditions}) and the field equations also imply a
special relation between the coupling constants $b_{i}$, which are constrained
to obey the equation
\begin{equation}
10b_{0}b_{2}=9b_{1}^{2}; \label{trecebis}%
\end{equation}
see \cite{nosotros} for a discussion. This can be phrasing as the solution
(\ref{g1})-(\ref{phi}), which exists in the theory defined at the point
(\ref{trecebis}) of the parameter space, represents a particle-like object
with \textit{charge} taking only three possible values: $-|q|$, $0$, or $+|q|$
(that is to say, $n$ in (\ref{phi}) takes the values $-|n|$, $0$, or $+|n|$).
However, the solution with $q>0$ is somehow patohological; consequently, here
we will mainly focus in the case $q\leq0$.

\section{Conserved charges}

Before going into the discussion of thermodynamics, let us compute the mass of
the black hole solution (\ref{g1})-(\ref{g2}). The most convenient way to do
this is resorting to the Hamiltonian formalism and working in the
minisuperspace approach.

One evaluates the action functional in an ansatz compatible with the
stationary configuration (\ref{g1}) and integrates in time in an interval ($t
$, $t+\Delta t$). The lapse function $N(r)$ in (\ref{g1}) enters as a Lagrange
multiplier whose coefficient, when varying the action, gives the variation of
the energy. Then, the method is varying the action and isolating the term
proportional to $N(r)$ taking care of boundary terms appropriately
-derivatives of $N(r)$ appear in the expression and one gets rid of them by
integrating by parts-. The variation of the action finally reads
\begin{equation}
\delta\mathcal{I}\approx-\dfrac{3\pi}{8G}\ N(\infty)\ \Delta t\ r^{2}\ \delta
f(r)\,, \label{trece}%
\end{equation}
with $\delta f(r)=-\delta m/r^{2}\,$. From this, one reads the boundary term
$\delta\mathcal{B}$ needed in the action to cancel the variation;\ and since
$N(r)$ is the Lagrange multiplier proportional to the conserved energy, one
reads from (\ref{trece}) the value of the mass $M$,
\begin{equation}
M=\dfrac{3\pi}{8G}m-M_{0}. \label{M1}%
\end{equation}
where $M_{0}$ is a constant to be determined. The interpretation of $M_{0}$ is
that of the energy of the background respect to which one measures the black
hole mass. A natural proposal for $M_{0}$ is, of course, this to be zero.
However, there are in principle other possibilities. For instance, another
choice for $M_{0}$ could be the minimum value of the parameter $3\pi m/(8G)$
for which the geometry exhibits a horizon. We will discuss this point latter,
when rederiving (\ref{M1}) with the Euclidean action method and obtaining
$M_{0}=0$. Mass (\ref{M1}) also agrees with what stated in \cite{nosotros}.

In contrast to $m$, parameter $q$ does not yield a conserved charge, but, up
to the sign $\varepsilon= -1,0,+1$, its value is fixed in terms of the
coupling constants $b_{i}$, which, in addition, obey a specific relation among
them for the black hole solution to exist.

\section{Black hole thermodynamics}

The Hawking temperature associated to black hole solution (\ref{g1}%
)-(\ref{g2}) is given by
\begin{equation}
T(r_{+})=\dfrac{1}{\pi\ell^{2}r_{+}^{4}}\left(  \frac{q \ell^{2}}{4}%
+\frac{\ell^{2}}{2}r_{+}^{3}+r_{+}^{5}\right)  \,, \label{T}%
\end{equation}
where we denoted $\ell^{2}=-\Lambda/6$.

In the case $n=0$, this expression reduces to that of Schwarzschild-AdS$_{5}$
black holes, presenting a minimum value $T^{(q=0)}_{\text{min}}$ for black
holes with horizon radius $r^{2}_{+}=\ell^{2}/2$. For negative values of $n$,
in contrast, the temperature as a function of $r_{+}$ may also present a
maximun $T^{(q<0)}_{\text{max}}$ at a smaller radius, and then the sign of the
specific heat becomes positive again at shorter radii. This implies that, for
negative $q$, small mass black holes exist for which the evaporation time
becomes infinite. Unlike GR in AdS, in this theory black holes exist with
arbitrarly low (positive) temperature -although it does not imply they are stable against
tunneling, as we will discuss.

Temperature (\ref{T}) has extrema when $r_{0}^{3}+2q-2r_{0}^{5}/\ell^{2}=0$.
Notice also that the temperature behaves like that of the asymptotically
Schwarzschild-AdS black hole for distance larger that $r_{+}\gtrsim|\ell|$
provided $|\ell|>|q|^{1/3}$. On the other hand, if $q<0$, for sufficiently
small black holes, $r_{+}<|q|^{1/3}$, the behavior differs substantially from
that of Schwarzschild black hole. This is consistent with the fact that the
presence of the hair parameter $q$ represents short distance corrections in
the theory. The behavior resembles that of asymptotically flat Schwarzschild
black hole only in the range between scales $|q|^{1/3}<r_{+}<|\ell|$.

In general, for both signs of $q$ and for certain range of temperatures, eight
different solutions in the theory with the same temperature may exist. That
is, apart from the three solutions of GR with $q=0$, namely the large and
small Schwarzschild-AdS$_{5}$ black holes and thermal AdS$_{5}$ space, there
exist three hairy black hole solutions with charge $q<0$ ($b_1\varepsilon<0$)
and two hairy black hole solutions with charge $q>0$ ($b_1\varepsilon>0$) that
all have the same temperature within the range $T_{\text{min}}^{(q>0)}%
<T<T_{\text{max}}^{(q<0)}$, where here $T_{\text{min}}^{(q>0)}$ refers to the
minimum of $T$ for the $q>0$ black hole. The temperature of black holes with
$q>0$ as a function of $r_{+}$ is qualitatively similar to that of the GR
($q=0$) solutions, although one always finds the relation
\begin{equation}
T^{(q>0)}(r_{+})>T^{(q=0)}(r_{+})>T^{(q<0)}(r_{+}). \label{jerarquia}%
\end{equation}
As said, in this paper we will only consider the case $q\leq0$, because $q>0 $
presents pathologies.

For negative values of $q$, other interesting phenomena appear. For example,
one of the new features that solution (\ref{g1})-(\ref{g2}) exhibits for $q<0
$ is the existence of an inner horizon at $r=r_{-}\leq r_{+}$. Moreover, there
exists a precise value of mass relative to the coupling constant,
$M_{\text{ext}}$, such that the solution is extremal, in the sense that
horizons $r_{-}$ and $r_{+}$ coincide in that case. $M_{\text{ext}}$
corresponds to the minimal value of the parameter $m$ for which the solution
exhibits a horizon. In this case, the double root of metric function $f(r)$
occurs at $r=r_{\,\text{ext}}$, with $m$ obeying the cubic equation%
\begin{equation}
r_{\text{ext}}^{3}-2mr_{\text{ext}}-\frac{5}{2}q=0. \label{re}%
\end{equation}
This determines the value of $M_{\text{ext}}$. The extremal radius
$r_{\,\text{ext}}$ also obeys the quintic equation%
\begin{equation}
r_{\text{ext}}^{5}+\frac{\ell^{2}}{2}r_{\text{ext}}^{3}+\frac{q\ell^{2}}{4}=0,
\label{W}%
\end{equation}
and, then, it immediately follows from (\ref{T}) that, as probably expected,
the temperature of the extremal solution vanishes, $T(r_{\text{ext}})=0$. The
near horizon geometry of the extremal solution is the product space
AdS$_{2}\times S^{3}$ with a curvature given by $r_{\text{ext}}^{-2}$.

\section{Euclidean path integral}

The fact that in this theory the sufficiently small hairy black holes have
positive specific heat suggests that phase transition may take place at short
distances. In fact, we will show that a rich phase diagram emerges, with
different geometries (with different values of $q$) dominating in different
ranges of temperature. To investigate the phase transition in the
semiclassical approximation, we resort to the Euclidean path integral
approach: In asymptotically AdS space, the problem can be thought of as one at
finite volume and, therefore, since one is concerned with a question at finite
temperature, the appropriate thermodynamical quantity to look at is Helmholtz
free energy $F$, \cite{HawkingPage}. In the semiclassical approximation, the
free energy is given in terms of $\mathcal{I}_{E}$ by the relation
\begin{equation}
F=-\beta^{-1}\log Z=\beta^{-1}\mathcal{I}_{E}, \label{F}%
\end{equation}
where $\beta=T^{-1}$ is the inverse of the Hawking temperature and, as usual,
is given by the period of the compactified Euclidean time $\tau=it$ required
for the real Euclidean section of the geometry not to develope a conical
deficit at $r=r_{+}$.
%Then, from condition (\ref{conditions}) one can solve
%for $b_{0}$ as a function of $q,$ $b_{1},$ and $b_{2}$, and then replace this
%into the Euclidean action $\mathcal{I}_{E}$ to finally compute $F$; see (\ref{bardo}) below.
It is worth mentioning that the computation of $F$ (see (\ref{bardo}) below)
involves a regularization of the infrared divergence, and this is achieved by
background substraction in the standard way, i.e. by substracting the
Euclidean action evaluated on the thermal AdS$_{5}$ configuration. In
particular, this amounts to match the (red-shifted) period $\beta$ at fixed radius
$r=r_{\text{IR}}$ and then take the limit $r_{\text{IR}}\rightarrow\infty.$

The mass can also be obtained from the Euclidean action as follows%
\begin{equation}
M=\frac{\partial\mathcal{I}_{E}}{\partial\beta}. \label{Me}%
\end{equation}
This yields%
\begin{equation}
M=\dfrac{3\pi}{8G}m=\dfrac{3\pi}{8G}\left(  r_{+}^{2}-\dfrac{q}{r_{+}}%
+\dfrac{r_{+}^{4}}{\ell^{2}}\right)  \,, \label{M2}%
\end{equation}
which reproduces the result obtained by the Hamiltonian formalism if setting
$M_{0}=0$ in (\ref{M1}).

It is worthwhile pointing out that the mass has a minimum when
\begin{equation}
r_{+}^{5}+\frac{\ell^{2}}{2}r_{+}^{3}+\frac{q\ell^{2}}{4}=0.
\end{equation}
This occurs for the value of $M$ of the extremal solution, namely
$M=M_{\text{ext}}$, cf. (\ref{W}). Solutions with $M<M_{\text{ext}}$ exhibit
no event horizon and, therefore, correspond to naked curvature singularities.
This could have suggested to set the reference background in (\ref{M1}) to be
that with $M_{0}=M_{\text{ext}}$ for not to have in the spectrum positive
energy states that represent naked singularities. This would have been similar
to what happens, for instance, in the case of the three-dimensional black hole
\cite{BTZ}, where the value of the mass parameter that corresponds to
AdS$_{3}$ space does not coincide with the minimum value of the black hole
mass. However, the Euclidean action computation (\ref{Me}) yields the result
(\ref{M2}) with $M_{0}=0$.

Black hole entropy is also obtained from the Euclidean action through the
relation%
\begin{equation}
S=\beta\frac{\partial\mathcal{I}_{E}}{\partial\beta}-\mathcal{I}_{E}.
\label{Se}%
\end{equation}
This yields
\begin{equation}
S=\dfrac{\pi^{2}}{2G}r_{+}^{3}+k\,, \label{S}%
\end{equation}
with $k$ given by
\begin{equation}
k=\frac{40\pi^{3}}{9}\varepsilon b_{0}\left(  -\frac{18}{5}\frac{b_{1}}{b_{0}%
}\right)  ^{5/2}=\frac{40\pi^{3}}{9}b_{0}n^{15}\,.
\end{equation}

Expression (\ref{S}) coincides with the result obtained by using the Wald
entropy formula \cite{Wald}.

Recalling that the volume of the $3$-sphere is $V=2\pi^{2}$ one verifies that,
up to the additive constant $k$, entropy (\ref{S}) satisfies the area law;
namely
\begin{equation}
S=\dfrac{A}{4G}+const.\, \label{A}%
\end{equation}
with $A=2\pi^{2}r_{+}^{3}$. Additive constants in the area law are usual
features in theories with higher-curvature couplings (see e.g. \cite{JM}-\cite{Anabalon}), and
conformal invariance of our model allows no other dependence. Understanding
the presence of $k$ in (\ref{S}) and its implicances would be actually
desirable. This probably demands a better understanding of the role of
boundary terms in this context.

It is worthwhile pointing out that the entropy obtained in (\ref{Se}%
)-(\ref{S}), together with (\ref{T}) and (\ref{M2}), fullfil the first law of
black hole thermodynamics; namely, one verifies
\begin{equation}
dM=T\ dS.
\end{equation}

Let us now study the free energy and the phase transitions between different
configurations at fixed temperature.

\section{Phase transition}

Then, one can compute the free energy $F=M-TS$. This can be written as%
\begin{equation}
F(r_{+})=-\dfrac{\pi}{8G\ell^{2}}r_{+}^{2}\left(  r_{+}^{2}-\ell^{2}%
+\frac{4q\ell^{2}}{r_{+}^{3}}\right)  -kT(r_{+}). \label{bardo}%
\end{equation}

In the case $q=0$ (i.e. $n=0$) this result reduces to that of GR, namely
$F=-\pi/(8G\ell^{2})r_{+}^{2}(r_{+}^{2}-\ell^{2})$, which in particular
manifestly exhibits the Hawking-Page transition \cite{HawkingPage} at
$r_{+}^{2}=\ell^{2}$. At large $r_{+}$, the difference between the free energy
of hairy ($q\neq0$) black holes and the free energy of the
Schwarzschild-AdS$_{5}$ ($q=0$) black holes goes like
\begin{equation}
\Delta F \sim - \dfrac{192 \pi^2 n^3 b_2 r_+}{3 \ell^2}+\mathcal{O}\left(\frac{1}{r_+}\right)
\label{follows}%
\end{equation}
This
implies that, for sufficiently large temperature, the hairy black holes with
$b_{2}n>0$ are thermodynamically favored.

Deciding which configuration dominates in each range of temperatures amounts
to compute the difference of the corresponding free energies. It is worthwhile
noticing that the comparison between the free energies of different solutions
has to be performed at fixed temperature $T$, which means to compare black
holes configurations that do not necessarily have the same radius $r_{+}$.
This is because the temperature as a function of the radius depends also on
the parameter $q$, so that two black holes with the three different values that $q$ may take (namely $0, \pm|q|$) and that
possess the same temperature necessarily have different radii (cf.
(\ref{jerarquia})).

\begin{figure}[h]
    \includegraphics[scale=0.3]{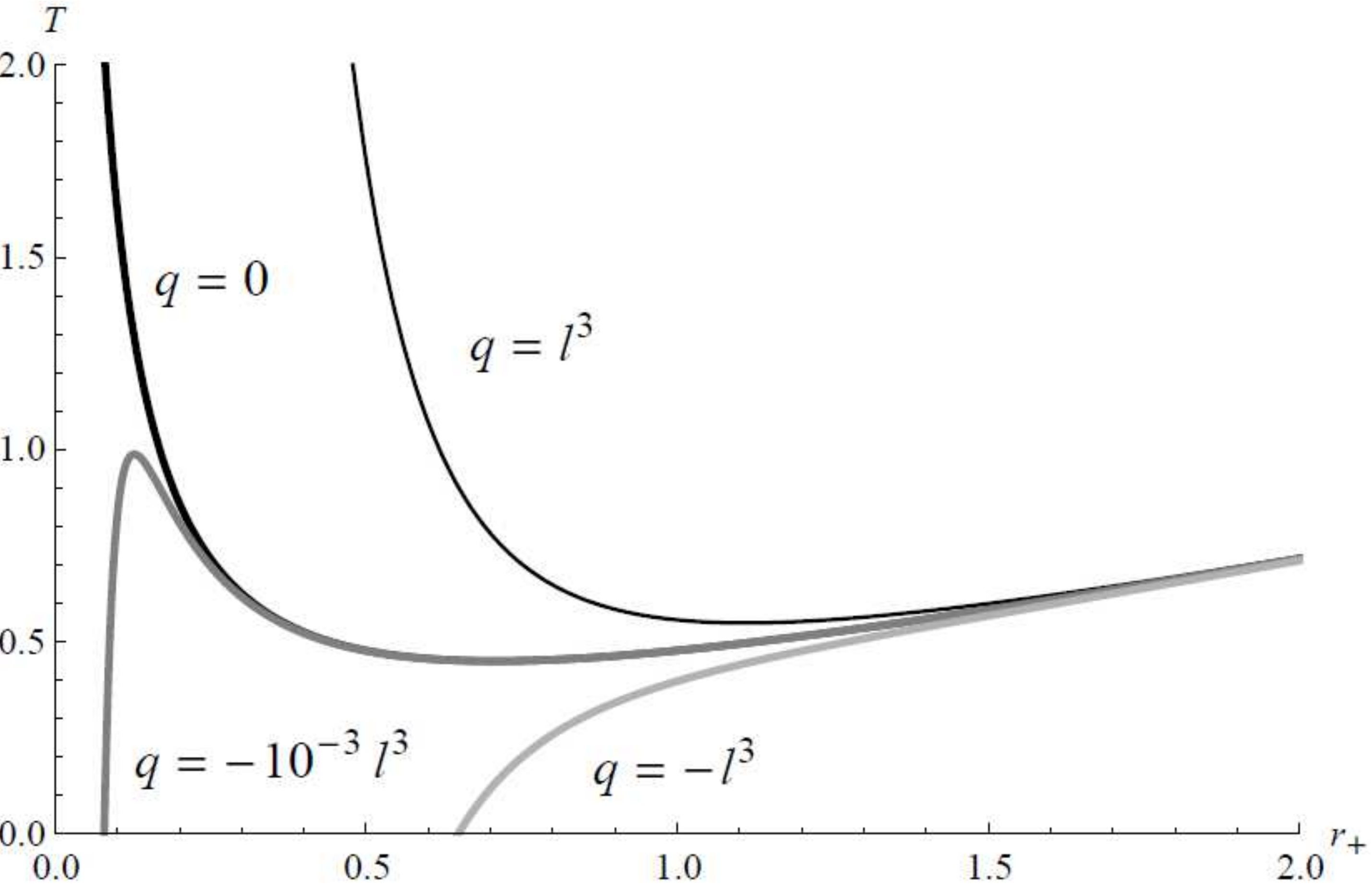}
    \caption{Temperature vs $r_{+}$ for different values of $q$}
    \label{Tvsrmas}
\end{figure}

The analysis of the main thermodynamical properties and the phase transitions
are summarized in Figures 1, 2, and 3. Figure 1\ shows the temperature
($T$)\ as a function of the (external) horizon radius ($r_{+}$) for different
values of $q$. The temperature curve of hairy black holes with $q_c<q<0$ exhibits
a local maximum ($q_c:=-\frac{3}{500}\sqrt{30}\ell^3\approx-0.0329\ell^3$). Besides, black holes with $q<0$, unlike those with $q=0$, may
have arbitrarily low temperature. So, there is a minimum value of temperature,
$T_{\text{min}}^{(q=0)}$, for which black holes with $q=0$ and with $q<0$ may
have the same free energy. Evaluating free energy (\ref{bardo}) for a black
hole with $q<0$ at temperature $T_{\text{min}}^{(q=0)}$, one finds a positive
value regardless the absolute value of $q$. This implies that thermal
AdS$_{5}$ is the configuration that dominates at such temperature. On the
other hand, at very large temperature, it follows from (\ref{follows}) that
-for negative values of $b_{2}$- the configuration with minimum free energy
corresponds to the $q<0$ black hole. Therefore, one concludes from this that phase transition takes place at
certain temperature $T_{c}>T_{\text{min}}^{(q=0)}$. This is summarized in
Figure 2, which shows the free energy ($F$)\ as a function of temperature
($T$) for the three different values of $|q|<|q_c|$. At large $T$ the black holes with
$q<0$ are thermodynamically favored, having the lowest free energy. Below
certain critical temperature, the thermal AdS$_{5}$ configuration ($F=0$) is
the dominant one. The segment connecting points A and B corresponds to the small
black holes\ with positive specific heat. The curve connecting points B and C
corresponds to black holes solutions with $q<0$ having temperatures between
the local minimum and local maximum of $T$. This analysis manifestly shows
that phase transition takes place at high temperature, being the hairy black
holes with negative \textit{charge} $q$ those that are favored.

Finally, for completeness, we also consider a case with $|q|>|q_c|$ which is depicted in Figure 3 and where the hairless case ($q=0$) has been included for reference. The curve on the left represents the black holes with $q<q_c\approx-0.0329\ell^3$. These black holes have positive specific heat and they are the dominant configuration under tunnelling. For positive $q>|q_c|$, the black holes lying on the curve connecting the point D to E have negative heat capacity, while those on the curve connecting the points E and F have positive heat capacity.

\begin{figure}[h]
    \includegraphics[scale=0.25]{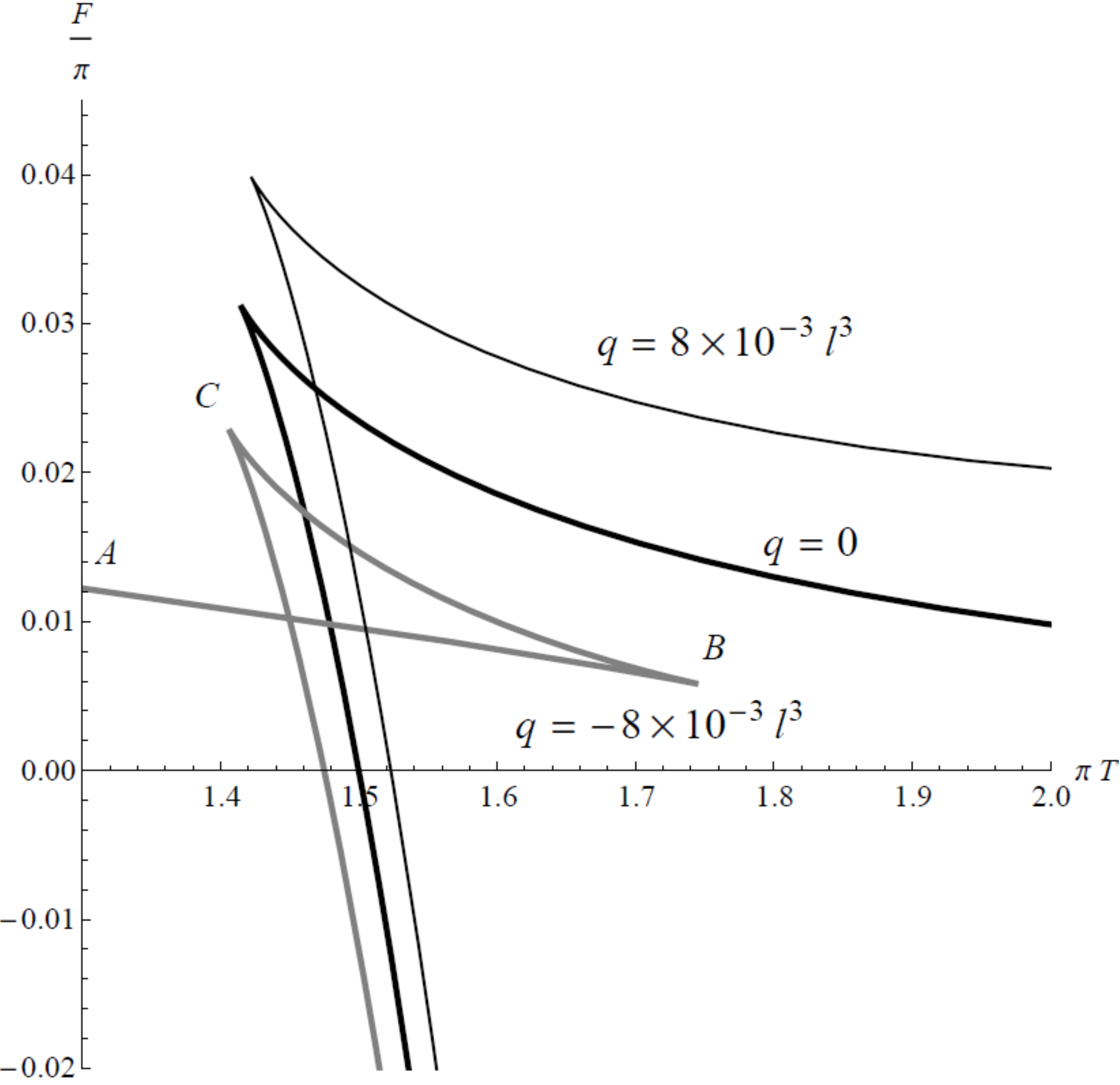}
    \caption{Free energy vs Temperature for different values of $|q|<|q_c|$}
    \label{FvsM}
\end{figure}

\begin{figure}[h]
    \includegraphics[scale=0.25]{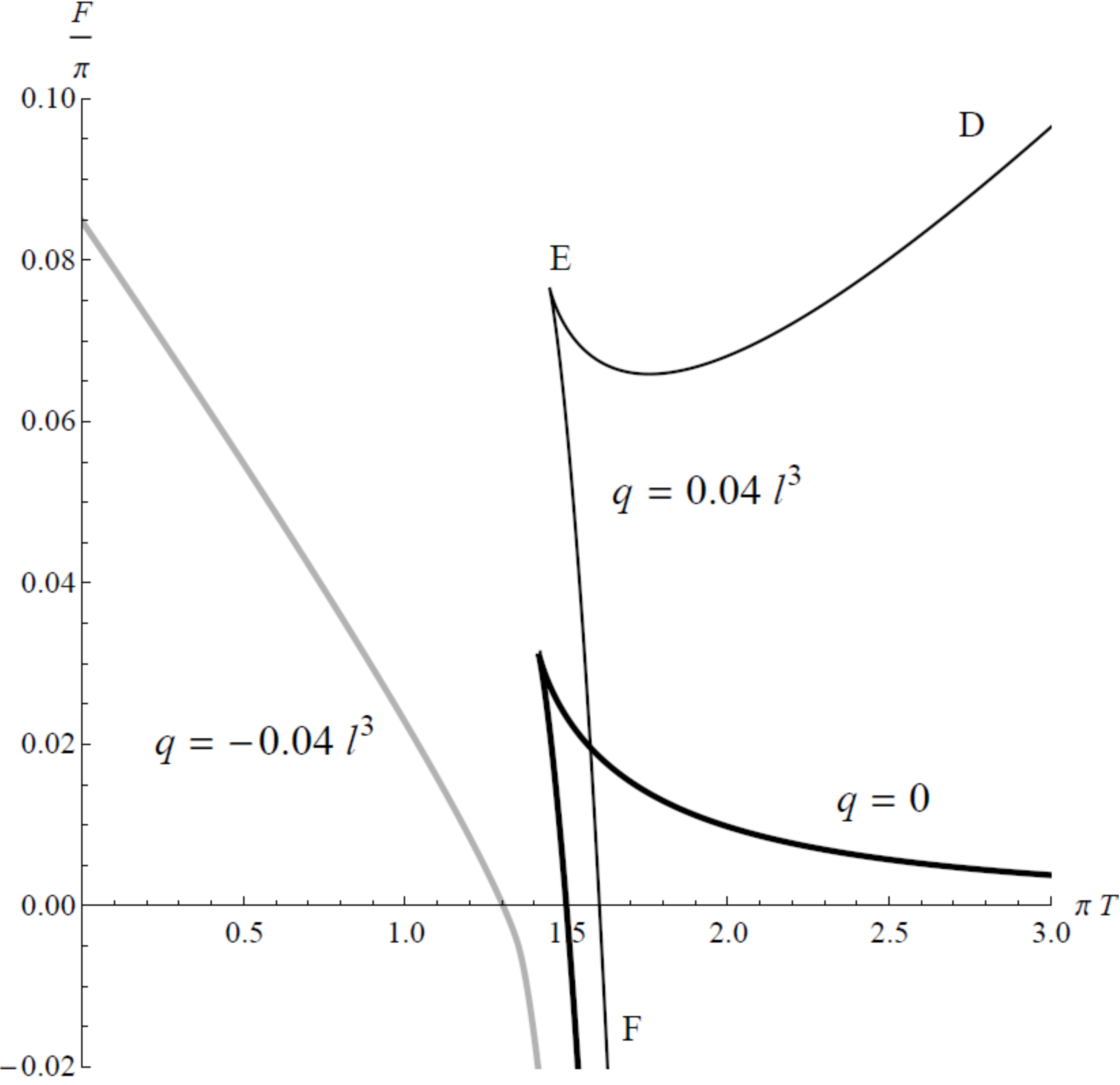}
    \caption{Free energy vs Temperature for different values of $|q|=0.04l^3>|q_c|$ (the hairless solution $q=0$ has been included for reference).}
    \label{FvsM}
\end{figure}

Figure 4, on the other hand, shows the temperature as a function of the mass
($M$) for different values of $q$ -corresponding to different choices of the
coupling constants in the matter Lagrangian-. Small black holes with $q<0$
presents positive specific heat. Through evaporation, these black holes yields
a zero-temperature remnant with non-vanishing mass. At first sight it may seem
that temperature of $q<0$ black holes can take negative values. However, such negative values
of $T$ are not physical as correspond to configurations that are not
accessible through evaporation.

\begin{figure}[h]
    \includegraphics[scale=0.25]{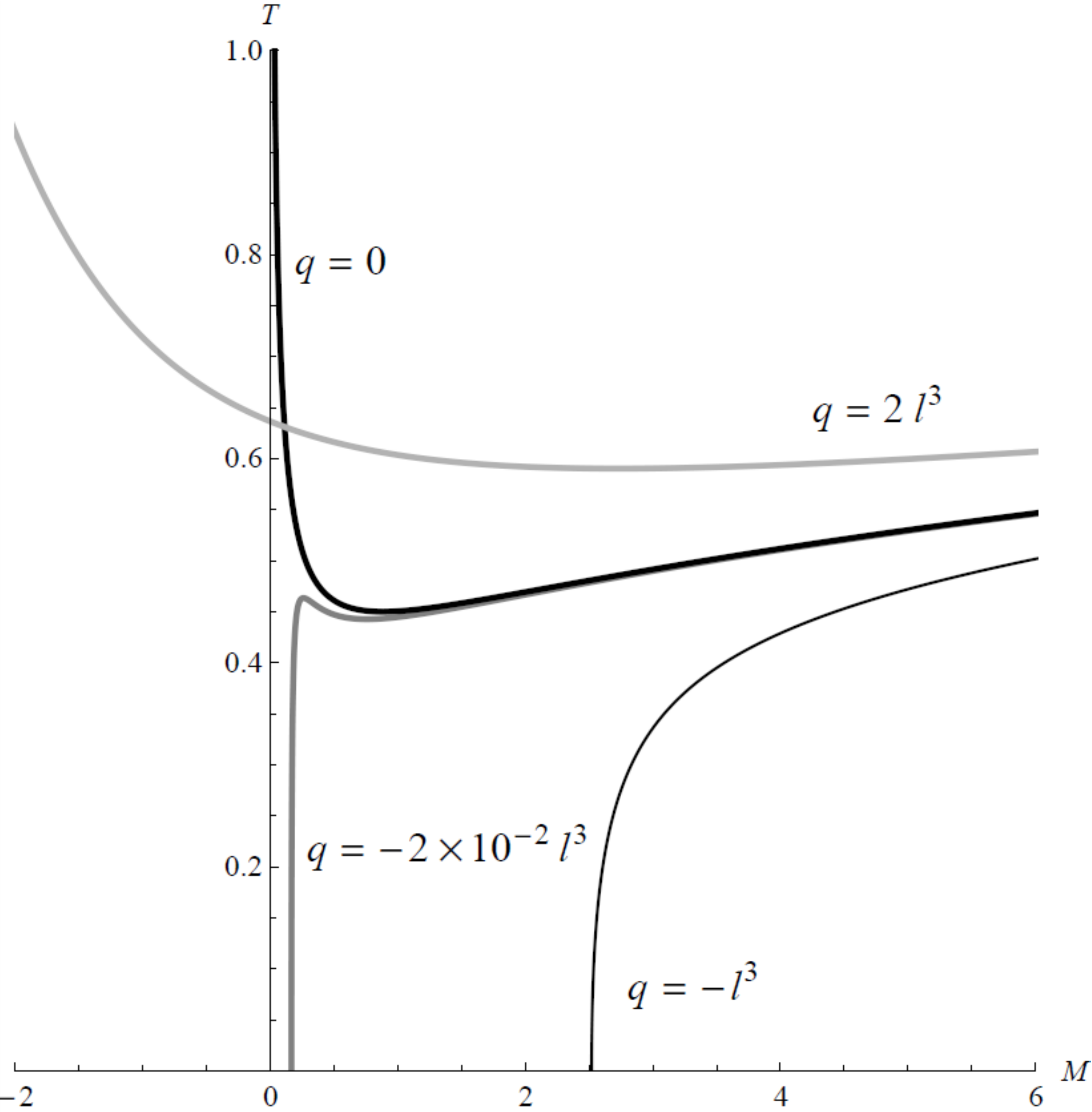}
    \caption{Temperature vs Mass for different values of $q$}
    \label{TvsM}
\end{figure}

\section{Conclusions}

In this paper we studied the thermodynamics of hairy black holes in
five-dimensional Anti-de Sitter space (AdS$_{5}$). These black holes are
solutions of General Relativity (GR)\ coupled to a conformally coupled scalar
field theory in five dimensions. This provides a simple setup in which phase transitions of hairy black holes in AdS can be solved explicitly, including backreaction. 

The hairy black hole solutions we consider exhibit a non-vanishing
scalar field configuration that is regular everywhere outside and on the
horizon, and behave at large distance in a way that respects the
asymptotically AdS$_{5}$ boundary conditions that are relevant for AdS/CFT.
The \textit{charge} of the scalar field may take three discrete values, $-|q|,$
$0,$ $+|q|$, each case exhibiting rather different properties. We studied the
thermodynamical properties of these solutions and we showed that, for certain
range of temperatures, black holes with non-vanishing scalar hair
configurations are thermodynamically favored with respect to the non-hairy
Schwarzschild-AdS$_{5}$ solution and thermal AdS$_{5}$ space. On the other
hand, below certain critical temperatre the theory undergoes Hawking-Page
transition and the dominant configuration happens to be thermal AdS$_{5}$. The
presence of the scalar conformal field also changes the thermodynamics of the
theory at short distance:\ In particular, we showed that, unlike what happens
in GR in absence of matter, small\ hairy black holes with arbitrarily low
temperature exist in AdS$_{5}$. These small black holes present a positive
specific heat, and as a consequence of that they become stable under Hawking
radiation, yielding a remnant with finite mass. However, they are not stable
under tunnelling since, for the range of temperatures such small black holes
exhibit, the free energy of thermal AdS$_{5}$ is lower. It would be interesting to explore whether this situation remains the same in higher dimensions or even when considering the black holes with hyperbolic horizons found in \cite{nosotros}.

\[
\]

This work has been supported by UBA, CONICET, and ANPCyT. This work has also been supported by FONDECYT Regular grant 1141073. The authors are
grateful to Mat\'{\i}as Leoni and Sourya Ray for discussions and collaboration
in this subject and thank as well \textit{La \'{U}ltima Frontera, Valdivia, Chile}, where part of these ideas were discussed.  A.G. thanks Universidad Austral de Chile for the hospitality. G.G. thanks Centro de Estudios Cient\'{\i}ficos CECs and
Pontificia Universidad Cat\'{o}lica de Valpara\'{\i}so for the hospitality
during his stays, where part of this work was done. J. O. thanks Universidad de Buenos Aires and the International Center for Theoretical Physics, ICTP, Trieste, Italy, were part of this project was carried out.

\end{document}